\documentclass[%
 reprint,
 amsmath,amssymb,
 aps,
]{revtex4-1}

\usepackage{graphicx}
\usepackage{dcolumn}
\usepackage{bm}

\begin{document}

\preprint{APS/123-QED}

\title{Energy Band Model Based on  Effective Mass }
\thanks{Thanks to:  Leon Altschul of SDL Ltd. and Vladimir Kleiner of ECI Ltd. for useful comments.}%
\author{Viktor Ariel}
\affiliation{%
 Department of Physical Electronics\\
 Tel Aviv University
}%

\begin{abstract}
In this work, we demonstrate an alternative method of deriving an isotropic energy band model using a one-dimensional definition of the effective mass and experimentally observed dependence of mass on energy.  We extend the effective mass definition to anti-particles and particles with zero rest mass.  We assume an often observed linear dependence of mass on energy and derive a generalized non-parabolic energy-momentum relation. The resulting non-parabolicity leads to velocity saturation at high particle energies. We apply the energy band model to free relativistic particles and carriers in solid state materials and obtain commonly used dispersion relations and experimentally confirmed effective masses.  We apply the model to zero rest mass particles in graphene and propose using the effective mass for photons.  Therefore, it appears that the new energy band model based on the effective mass can be applied to  relativistic particles and carriers in solid state materials. \\
\end{abstract}

\pacs{Valid PACS appear here}
\maketitle


\section{\label{sec:level1}Introduction\protect\\}  
In solid state materials, the energy-momentum dispersion relation $E(p)$ is traditionally derived by associating between particles and wave packets and then solving the wave equation  \cite{Harrison}, \cite{Seeger}. In this work, we demonstrate an alternative procedure suitable for isotropic materials. We propose a theoretical definition of the particle effective mass based on  \cite{ArielPaper} and  \cite{ArielArxiv1}. Then,  we use a known dependence of mass on energy, which can be obtained for example  from cyclotron resonance measurements in semiconductors \cite{Seeger}. Finally, we integrate the mass expression over the allowed energy range in order to obtain the dispersion relation.
As an example, we assume a linear approximation of mass as a function of energy, $m(E)$, which is often observed  experimentally for relativistic particles and  in solid state materials  \cite{ArielArxiv2}, \cite{Zawadzki}, \cite{Castro}. Finally, we demonstrate that the definition of the effective mass and the resulting dispersion relation are suitable for description of free relativistic particles,  carriers in solid state materials, and massless particles such as carriers in graphene and photons.

\section{Definition of the  Effective Mass } 

We begin with the  basic definition of the particle effective mass as presented in \cite{ArielArxiv1} and \cite{ArielArxiv2} where we identified particles with wave packets.  This one-dimensional approach can be used for any isotropic energy-momentum relation $E(p)$ \cite{ArielArxiv1}.

\begin{figure}
\includegraphics{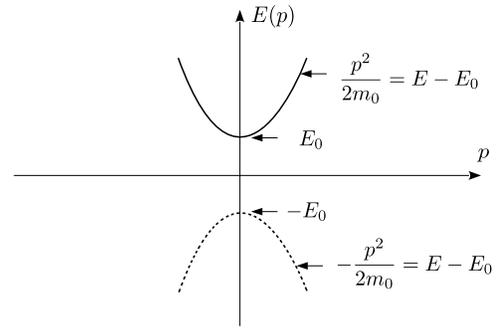}
\caption{\label{fig:energy-parab} Simple parabolic energy band model in semiconductors.}
\end{figure}

In Fig.~\ref{fig:energy-parab}, 
we show an example of a simple parabolic energy band model typically used in direct-gap semiconductors. This model is usually obtained from a solution of the wave equation and results in two energy bands: one band  for particles  with positive energy, similar to electrons in semiconductors, and another  for 
anti-particles with negative energy, similar to light holes. Assume that the dispersion relation $E(p)$ achieves a minimum with the energy $E_0$ for the  particle momentum $p=0$, also assume that  $E(p)$ is symmetrical with respect to $E=0$ and $p=0$.

The kinetic energy for parabolic electrons and holes can be defined  respectively as

\begin{equation}
 E-E_0 = \frac {p^2} {2m_0} , \ \ \  E-E_0 = -\frac {p^2} {2m_0} \ .
\label{eq:kinetic_energy}
\end{equation}

Since the electron is at rest when $p=0$, we can define $E=E_0$ as  the electron rest energy and the electron mass at the energy minimum, $m=m_0$, as the electron rest mass. Clearly for electrons $E \ge E_0$ while for holes $E\le ~E_0$. 

Usually \cite{Seeger}, \cite{ArielArxiv1},  the one-dimensional group velocity of the wave packet is defined  as,
\begin{equation}
 v_{g} =   \frac {\partial E} { \partial  p} \ .
\label{eq:group}
\end{equation}

\begin{figure}
\includegraphics{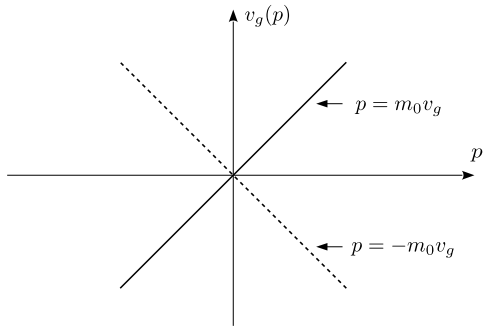}
\caption{\label{fig:group-parab} Relationship between the momentum and group velocity for parabolic electrons (solid line) and holes (dashed line) in semiconductors.}
\end{figure}

In Fig. \ref{fig:group-parab}, 
we show a relationship between particle group velocity and momentum for parabolic electrons and holes obtained from  (\ref{eq:kinetic_energy}) and (\ref{eq:group}).

Based on the semi-classical definition of the
particle momentum, the effective mass appears as a proportionality factor between the particle momentum and the group velocity of the corresponding wave packet. 

However, there appears a complication with the sign of the effective mass for holes since the momentum is in the opposite direction from the group velocity as can be seen in Fig.~\ref{fig:group-parab}. 

For electrons we obtain as expected,  $p=m_0 v_g$, while the hole momentum is a negative function of the hole group velocity, $p=-m_0 v_g$. This can be explained by the fact that the impact of moving holes is actually due to the electrons moving in the opposite direction \cite{Harrison}.

There is compelling experimental evidence that the effective mass should have a positive value for both electrons and holes.
This can be deduced from the fact that electrons and holes rotate in opposite directions under the influence of the magnetic field. The implication is that the ratio of the electric charge to the effective mass has the opposite sign  for electrons and holes. Electrons have a negative charge and holes behave like electrically positive charges, as confirmed for example by their electrical attraction to electrons. Consequently, both electrons and holes should have positive effective mass.

Therefore, we propose using the absolute values of the particle momentum and group velocity in order to define the effective mass for both particles and anti-particles with the help of (\ref{eq:group}),

\begin{equation}
 m(E) \equiv  \frac {|p|} {|v_g|} \equiv \frac {|p|} {|\partial E/ \partial p|} \ .
\label{eq:mass}
\end{equation}

Note that the effective mass defined by (\ref{eq:mass}) is generally energy dependent and by definition always positive for both electrons and holes.

\begin{figure}
\includegraphics{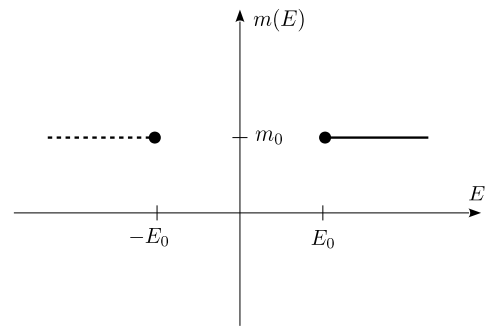}
\caption{\label{fig:mass-parab} Effective mass as a function of energy for parabolic energy bands.}
\end{figure}

In parabolic materials, the effective mass is a constant, $m_0$, independent of energy as can be seen from (\ref{eq:kinetic_energy}) and (\ref{eq:mass}) and demonstrated  in  Fig.~\ref{fig:mass-parab}. 
Since there are no particles for $|E| < E_0$, the mass is not defined for that energy range.

Therefore based on the semi-classical definition of the particle momentum, we can define the  effective mass (\ref{eq:mass})  using the absolute values of the particle momentum and group velocity  which gives positive values of the effective mass for both particles and anti-particles.

\section{Linear effective mass approximation} 

One of the main disadvantages of the parabolic energy model is the lack of velocity saturation at high particle energies, which is observed experimentally for both relativistic particles and semiconductors. Assume that in the vicinity of $E=E_0$, the effective mass can be approximated as a linear function of energy as observed  from cyclotron resonance measurements in semiconductors  \cite{Zawadzki},  \cite{Castro} and from $|E|=mc^2$ for relativistic particles. Therefore, we assume the following  linear approximation for the effective mass,

\begin{equation}
m(E)\simeq \frac { |E|} {v_0^2}\ ,
\label{eq:mass-linear-e0}
\end{equation}
where $v_0$ is a constant velocity that can be identified as the particle saturation velocity, as we show later.

At the bottom of the electron energy band, $E=E_0$ and $m=m_0$, and we obtain  from (\ref{eq:mass-linear-e0}),
\begin{equation}
E_0 = m_0 v_0^2\ .
\label{eq:rest-energy}
\end{equation}
A similar result is obtained for anti-particles from symmetry.
Assuming  that there are no particles present for energies $|E|~<~E_0$, we can define the forbidden energy gap as 
\begin{equation}
E_G=2E_0 = 2 m_0 v_0^2\ ,
\label{eq:band-gap}
\end{equation}
which demonstrates dependence between the particle bandgap, rest mass, and saturation velocity.

\begin{figure}
\includegraphics{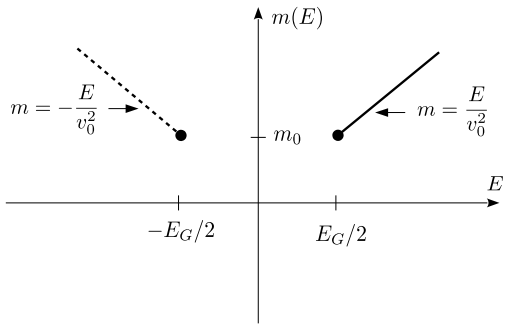}
\caption{\label{fig:mass-energy} Effective mass as a function of energy for particles and anti-particles.}
\end{figure}

\begin{figure}
\includegraphics{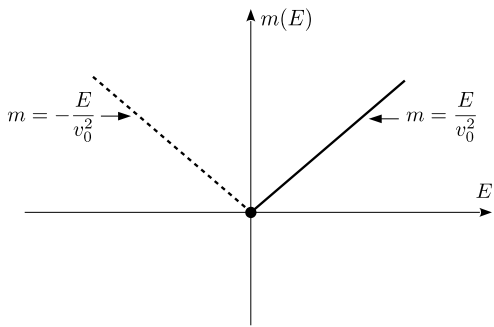}
\caption{\label{fig:mass-energy-zero} Effective mass as a function of energy for zero rest mass particles.}
\end{figure}

Then, we obtain for the particle effective mass from  (\ref{eq:mass-linear-e0}) and (\ref{eq:band-gap}),
\begin{equation}
m(E) =\frac { |E|} {v_0^2} = \frac {2 m_0 |E|} {E_G} \ .
\label{eq:mass-linear}
\end{equation}
In Fig.~\ref{fig:mass-energy}, we show mass-energy dependence for particles and anti-particles based on (\ref{eq:mass-linear}). 

For particles with zero rest mass,  we assume that $m_0~\sim ~0$ and consequently from (\ref{eq:band-gap}) $E_G\sim 0$. However,  $v_0\ne0$, and we still can define the effective mass from (\ref{eq:mass-linear}) as,
\begin{equation}
m(E)\simeq  \frac { |E|} {v_0^2}\ .
\label{eq:mass-linear-zero}
\end{equation}
It appears from (\ref{eq:mass-linear-zero}) that  for zero rest mass particles, the dependence of mass on energy is similar to regular particles only with an assumption that  $m_0$ and $E_G$ are much smaller than the particle energy, 
$|E| \gg m_0 v_0^2$ or equivalently  $|E| \gg E_G/2$.  The resulting $m(E)$ is presented in Fig.~\ref{fig:mass-energy-zero}. 

Therefore, we can model the effective mass of regular particles, anti-particles, and  zero rest mass particles  using the same linear effective mass approximation (\ref{eq:mass-linear}).

\section{Dispersion relation based on the effective mass } 

Assume that we can experimentally determine the dependence of the effective mass  on energy in (\ref{eq:mass}), for example using cyclotron resonance in semiconductors \cite{ArielArxiv2}. Then, we can obtain the dispersion relation $E(p)$ by  integrating (\ref{eq:mass})  over allowed energy values, which for one-dimensional isotropic case leads to
\begin{equation}
 p^2(E)  =  \int^{E}  2m(E^{'})\  d E^{'}\ .
\label{eq:momentum-energy}
\end{equation}

Substituting the linear mass  (\ref{eq:mass-linear}) into (\ref{eq:momentum-energy})   results in
 
\begin{equation}
p^2=\int_{E_G/2}^E \frac {4 m_0 |E|} {E_G^2} \  d E\ .
\label{eq:momentum-integral}
\end{equation}

Integrating equation (\ref{eq:momentum-integral})
we obtain for particles
\begin{equation}
p^2= \frac {2 m_0 E^2} {E_G} - \frac {m_0 E_G} {2} =  \frac {E^2} {v_0^2}-m_0^2 v_0^2\ .
\label{eq:momentum-general}
\end{equation}
An identical expression results for anti-particles.
Close to the bottom of the band, where $p\sim 0$, we can approximate for particle energy from (\ref{eq:momentum-general})
\begin{equation}
\frac {p^2} {2 m_0}\simeq   E- m_0 v_0^2  \ .
\label{eq:momentum-lowe}
\end{equation}
Thus, a parabolic dispersion relation is preserved at the bottom of the energy band.
\begin{figure}
\includegraphics{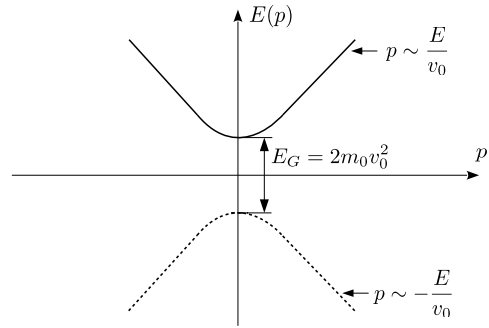}
\caption{\label{fig:energy-momentum}  Two-band non-parabolic energy model resulting from the linear effective mass.}
\end{figure}
Assuming that for high energy particles,  $ |E |\gg m_0 v_0^2$ or equivalently $|E| \gg E_G/2$, we can approximate  from  (\ref{eq:momentum-general}) for high energies
\begin{equation}
p \sim \pm  \frac {E} { v_0}\ .
\label{eq:momentum-highe}
\end{equation}

For the group velocity of high energy particles we obtain from (\ref{eq:group}) and  (\ref{eq:momentum-highe})

\begin{equation}
v_{g} \sim  v_0 .
\label{eq:velocity-highe}
\end{equation}

Therefore, the particle velocity approaches a maximum  saturation velocity $v_0$ for high energies. The velocity saturation is a direct consequence of the linear mass-energy dependence in (\ref{eq:mass-linear}). The theoretical velocity saturation is important  because it is observed experimentally for high energy particles. We can calculate $v_0$ from (\ref{eq:band-gap})
\begin{equation}
v_0 = \sqrt {\frac { E_G} {2m_0}}  \ .
\label{eq:sat-velocity}
\end{equation}

The resulting two-band energy model obtained from   (\ref{eq:momentum-general})  and  (\ref{eq:sat-velocity})  is presented in Fig.~\ref{fig:energy-momentum}. 
It appears that the main parameters of the model $E_G$, $m_0$, and $v_0$ can be easily measured experimentally.

\begin{figure}
\includegraphics{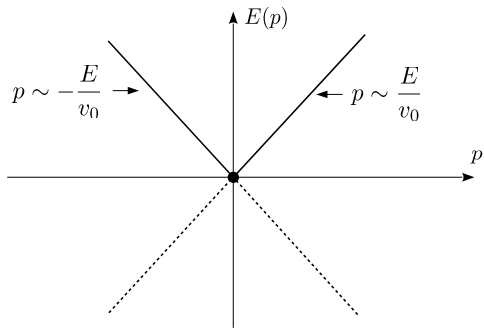}
\caption{\label{fig:energy-momentum-zero}  Two-band energy model for particles with zero rest mass.}
\end{figure}

The dispersion relation for zero rest mass particles results from  (\ref{eq:mass}),  (\ref{eq:mass-linear-zero}), and (\ref{eq:momentum-general})  by assuming $ |E |\gg m_0 v_0^2$ leading to,
\begin{equation}
p=\pm \frac {E} {v_0} \ .
\label{eq:momentum-energy-zero}
\end{equation}
This in-turn leads to the group velocity of zero rest mass particles from (\ref{eq:momentum-energy-zero})
\begin{equation}
v_g  =  v_0 .
\label{eq:velocity-zero}
\end{equation}
As expected, zero rest mass particles always propagate with maximum saturation velocity and have linear dependence of the particle 
momentum on energy similar to high energy massive particles.

It appears from 
Fig.~\ref{fig:energy-momentum-zero} 
that the energy band model for zero rest mass particles (\ref{eq:momentum-energy-zero}) can be considered an extension of the regular two-band model (\ref{eq:momentum-general}) for high energy particles. 

Thus, we were able to demonstrate a simple derivation of the particle dispersion relation based on the  definition of the effective mass and linear dependence of mass on energy.

\section{Application to relativistic particles and semiconductors} 
We can apply the general dispersion relation (\ref{eq:momentum-general})  to free relativistic particles and carriers in solid state materials.

For  relativistic particles, the saturation velocity is equal to the speed of light $v_0 =c \ $ and  $E_G=  2 m_0 c^2$. Then,  from (\ref{eq:mass-linear})  we obtain the usual relativistic mass-energy dependence,
\begin{equation}
m=\frac {|E|} {c^2} \ .
\label{eq:mass-Einstein}
\end{equation}
Applying the general dispersion relation (\ref{eq:momentum-general}) to free relativistic particles  (\ref{eq:mass-Einstein}),  we obtain as expected: 

\begin{equation}
p^2 = \frac {E^2} {c^2} - m_0^2 c^2 \  .
\label{eq:momentum-relativistic}
\end{equation}
Note that (\ref{eq:momentum-relativistic}) was derived from the linear mass assumption and not from the traditional relativistic concepts.

Next, we apply (\ref{eq:momentum-general}) to electrons in solid state materials where the energy reference is usually chosen at the bottom of the conduction band. The energy 
$E$ is then identified as the electron kinetic energy.  Under these assumptions, $m=m_0$ for $E=0$, which leads  from (\ref{eq:mass-linear})  to
\begin{equation}
m\simeq m_0 \left(1 + 2 \frac { E} {E_G}\right)\ .
\label{eq:mass-semi}
\end{equation}

Using  (\ref{eq:momentum-general}), we derive the energy band model for electrons in solid state materials, 
\begin{equation}
\frac {p^2} {2m_0}\simeq E+ \frac {E^2} { E_G}\  .
\label{eq:momentum-semi}
\end{equation}
The above approximation (\ref{eq:momentum-semi}) is similar to the simplified two-band Kane model  \cite{Harrison}, \cite{Seeger}. This model is usually derived by a complex solution of the wave equation. It is verified experimentally and commonly used in non-parabolic semiconductors such as  HgCdTe  and InSb \cite{ArielPaper}, \cite{Zawadzki}. 

Therefore,  the resulting dispersion relations for  free relativistic particles (\ref{eq:momentum-relativistic}) and carriers in solid state materials (\ref{eq:momentum-semi}) are equivalent since both are obtained from the same general relationship (\ref{eq:momentum-general}) by a suitable choice of the saturation velocity and energy reference.

Next, we apply the effective mass model to zero rest mass particles.
Carriers in graphene can be described as  zero rest mass fermions with the saturation velocity equal to the Fermi velocity $v_0=v_F$ \cite{Castro}. Then, for carriers in graphene we obtain from (\ref{eq:momentum-energy-zero})
\begin{equation}
p\simeq  \pm \frac {E} {v_F} \ , \ \  v_g\simeq v_F, \ m\simeq \frac {|E|} {v_F^2}\ ,
\label{eq:mass-graphene}
\end{equation}
which is confirmed experimentally from cyclotron resonance measurements \cite{Castro}.

While the carrier effective mass in graphene is observed experimentally, the concept of the effective mass is not commonly applied to photons. Nevertheless,  by using  $v_0 =c$ and $E= \hbar \omega$  we  can define the photon effective mass (\ref{eq:mass}), 
\begin{equation}
 m\simeq \frac {\hbar \omega} {c^2} \ .
\label{eq:mass-photon}
\end{equation}

Thus, we demonstrate that a simple energy band model derived using the effective mass definition and linear dependence of mass on energy is the same for relativistic particles, carriers in semiconductors, and zero rest mass particles.

\section{Conclusions}

In this work, we extend the one-dimensional effective mass definition to particles, anti-particles, and zero rest mass particles by using the absolute values of the particle group velocity and momentum. We demonstrated that we can derive an energy band model based on the effective mass definition and using experimentally observed linear dependence of mass on energy. The resulting energy bands are non-parabolic and are similar to the electron and light hole bands in non-parabolic semiconductors. The energy non-parabolicity is a direct consequence of the linear dependence of the effective mass on energy and leads to carrier velocity saturation at high energies. The model for zero rest mass particles can be considered a high energy extension of the regular energy model. We demonstrated that the resulting expressions  for  free relativistic particles and carriers in semiconductors are equivalent and only depend on the choice of the reference energy. We applied the dispersion relation to zero rest mass particles such as  carriers in graphene and  photons. In conclusion, it seems that the proposed energy band model derived from  the effective mass  is compatible with the traditional dispersion relations and can be useful for description of relativistic particles and carriers in solid state materials.

\pagebreak


\begin{thebibliography}{9}

\bibitem{Harrison}
W. A.  Harrison, 
\emph{Solid State Theory}
(McGraw-Hill, 1970).

\bibitem{Seeger}
K. Seeger, 
\emph{Semiconductor Physics}
(Springer-Verlag, 1985).

\bibitem{ArielPaper}
V. Ariel, A. Fraenkel, E. Finkman, J. Appl. Phys. {\bf 71}, 4382 (1992).

\bibitem{ArielArxiv1}
V. Ariel, arXiv:1205.3995v1 [physics.gen-ph], (2012).

\bibitem{ArielArxiv2}
V. Ariel, A. Natan,  	arXiv:1206.6100v1 [physics.gen-ph], (2012).


\bibitem{Zawadzki}
W. Zawadzki, S. Klahn, U. Merkt, Phys. Rev. Lett.  {\bf 55}, 983 (1985).



\bibitem{Castro}
A. H. Castro Neto, F. Guinea,  N. M. R. Peres, K. S. Novoselov, A. K. Geim,
“The electronic properties of graphene”, Rev.  Mod. Phys., {\bf 81}, 109 (2009).


\end{thebibliography}
\end{document}